\documentstyle[aps,epsf,prb]{revtex}
\begin{document}
\twocolumn
\def\bfr{{\bf r}}
\def\bfk{{\bf k}}
\def\bfq{{\bf q}}
\def\rmD{{\rm D}}
\def\rmkin{{\rm kin}}
\def\rmex{{\rm ex}}
\def\modv{\mid {\bf v} \mid}

\title{Local-energy density functionals for an N-dimensional \\ electronic 
system in a magnetic field}

\author{B.P.~van Zyl}
\address{Department of Physics, Queen's University,\\ Kingston,
Ontario, Canada K7L 3N6}
\date{\today}

\maketitle

\begin{abstract}
We present a general approach for the construction of the exact
local-energy density functionals for a uniform
N-dimensional electronic system in a magnetic field.  
For {\em arbitrary} dimension, we obtain explicit expressions for the
matter, kinetic, and exchange density functionals.
In the zero-field limit, we recover the usual N-dimensional Thomas-Fermi
theory.   As an application of our results, we develop a
current-density-functional theory, in the spirit of the
Thomas-Fermi-Dirac approximation, for an inhomogeneous many-electron system
in a magnetic field. 

\end{abstract}

\section{INTRODUCTION}
\label{intro}

It is now well established that a density-functional approach 
to the ground state and dynamical properties of many-electron systems
is a viable alternative to the usual wavefunction approach
involving the solution of the single-particle Schr\"odinger equation.
The proof by Vignale and Rasolt \cite{vignale} 
(VR) that the scalar and vector potentials characterizing a
many-electron system are uniquely determined by the single-particle electron
and current densities (the so-called current-density-functional theory CDFT)
has placed density-functional theory on a 
rigorous footing for the description of many-electron systems involving 
electric and magnetic fields.

In this work, we present a general approach for the construction of the
local-energy density functionals for an N-dimensional electronic system
in a homogeneous magnetic field.  Unlike the Kohn-Sham single-particle scheme, 
in which only the exchange-correlation functional is unknown, this statistical
formalism requires the knowledge of both the kinetic and exchange energy
functionals in the presence of electric and magnetic fields.
Our approach is to consider the exact canonical density matrix for a
noninteracting N-dimensional electron gas in a homogeneous magnetic field.
The single-particle density matrix is obtained via the inverse
Laplace transform of the canonical density matrix, as originally formulated
by March and Murray \cite{march}.  The diagonal component of the single-particle
density matrix yields
the exact density distribution of the system.  The exact kinetic and exchange
energy-density functionals may also be explicitly evaluated from the 
single-particle density matrix.  Although these functionals are obtained for 
a homogeneous system, an application
of the local-density approximation (LDA), in the spirit of 
Thomas-Fermi-Dirac theory \cite{dreizler}
(TFD), allows these functionals to be also used in inhomogeneous systems.  

Our paper is organized as follows.  In Sec. \ref{density-matrix}, we define
the N-dimensional canonical density matrix for a noninteracting electron gas
in a homogeneous magnetic field, and discuss some of the mathematical 
properties of the associated N-dimensional single-particle density matrix.
Then, in Sec. \ref{functionals}, we consider the evaluation of the 
kinetic and exchange energy-density functionals for arbitrary
dimension and magnetic field strength.   Explicit forms for the kinetic
and exchange energy-densities are obtained for three-and-two-dimensional
systems.  In Sec. \ref{cdft}, we present an outline for the
construction of a current-density-functional theory
based on the TFD approximation in two-and-three-dimensions, making use of the
density functionals obtained in Sec. \ref{functionals}.  Finally, in 
Sec. \ref{conclusions} we offer our concluding remarks.

\section{DENSITY MATRIX OF A N-DIMENSIONAL HOMOGENEOUS ELECTRON GAS IN A UNIFORM
MAGNETIC FIELD}
\label{density-matrix}

\begin{figure}
\begin{center}
\leavevmode
\hbox{%
\epsfxsize=3in
\epsffile{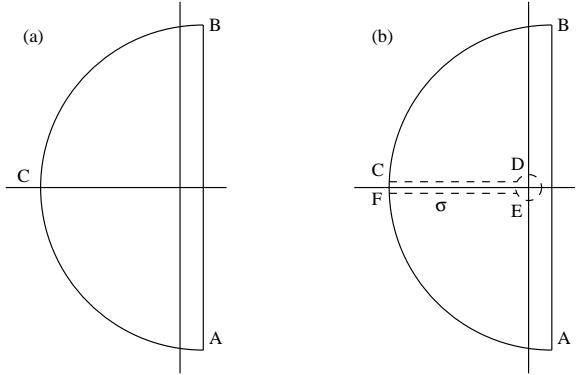}}
\end{center}
\caption{(a) The contour to be used when $d$ is integral, and (b) when $d$ is
half-integral.}
\label{contour}
\end{figure}

Long ago, Sondheimer and Wilson \cite{sondheimer}
 considered the evaluation of the diamagnetism
of free electrons.  The central tool in their calculation was the
determination of the canonical density matrix, which in 3-dimensions (3D)
is given by (including spin degeneracy)

\begin{eqnarray}
C(\bfr,\bfr';\beta) &=& \left( \frac{m}{2\pi\hbar^2\beta}\right)^{3/2}
\frac{\hbar\omega\beta}{{\rm sinh}(\hbar\omega\beta/2)} \times \nonumber \\
&{\rm exp}&\left\{-\frac{m}{2\hbar^2\beta}[ i\hbar\omega\beta(x'y-y'x) 
\right. \nonumber \\ &+&
\left. \frac{\hbar\beta}{2}{\rm coth}(\hbar\omega\beta/2)((x-x')^2+
(y-y')^2) \right.\nonumber \\
&+& \left. (z-z')^2 ]\right\}~,
\end{eqnarray}
where the symmetric gauge \cite{note1} has been used and $\omega = eB/mc$.
For purely formal reasons, we extend the definition of $C(\bfr,\bfr';\beta)$
to N-dimensions in the following way

\begin{eqnarray}
C(\bfr,\bfr';\beta) &=& \left( \frac{m}{2\pi\hbar^2\beta}\right)^{{\rm N}/2}
\frac{\hbar\omega\beta}{{\rm sinh}(\hbar\omega\beta/2)} \times \nonumber \\
&{\rm exp}&\left\{-\frac{m}{2\hbar^2\beta}[i\hbar\omega\beta(x'y-y'x)
\right. \nonumber \\ &+&
\left.\frac{\hbar\omega\beta}{2}{\rm coth}(\hbar\omega\beta/2)((x-x')^2 
(y-y')^2)]\right\}  \nonumber \\
&\times&\Pi_{i=1}^{{\rm N}-2}{\rm exp}\left(-\frac{m}{2\hbar^2\beta}(z_i - z_i')^2
\right)~,
\label{canonical}
\end{eqnarray}
with $({\rm N}=2, 3,...)$. 
The single-particle density matrix in N-dimensions is then obtained via 
the usual inverse Laplace transform relation \cite{gradshteyn}

\begin{eqnarray}
\rho(\bfr,\bfr') &=& \int_{\epsilon - i\infty}^{\epsilon + i\infty}
\frac{d\beta}{2\pi i}\frac{e^{\beta\varepsilon_f}}{\beta} 
C(\bfr,\bfr';\beta)~,
\label{den-matrix}
\end{eqnarray}
with $\varepsilon_f$ being the Fermi energy,
and the contour chosen to be closed in the left-half of the $\beta$-plane
(see Fig. \ref{contour}).  We defer an exact treatment of 
Eq. (\ref{den-matrix}) until the next section.

The electron number density $\rho$ (spin-averaged, zero temperature)
is immediately given by taking the
trace of the density matrix in Eq. (\ref{den-matrix}), viz., 

\begin{eqnarray}
\rho^{(d)}(\gamma) &=& 2\left(\frac{m\omega}{4\pi\hbar}\right)^{d} 
{\cal F}^{(d)}(\gamma)~,
\label{rhod}
\end{eqnarray}
with 
\begin{eqnarray}
{\cal F}^{(d)}(\gamma) \equiv \int_{\epsilon - i\infty}^{\epsilon + i\infty}
\frac{ds}{2\pi i}\frac{e^{2\gamma s}}
{s^{d}{\rm sinh}(s)}~,
\label{trace}
\end{eqnarray}
and for notational convenience, we have defined $s \equiv \hbar\omega\beta/2$,
$\gamma \equiv \varepsilon_f/(\hbar\omega)$, and $d={\rm N}/2$.
Depending on the dimension of the space, $d$ can be either integral or
half-integral.  

In the case that $d$ is half-integral (i.e., odd-N), the
inverse Laplace transform defined by ${\cal F}^{(d)}$ {\em always} contains
a branch point in the complex $s$-plane at $s=0$.  
Therefore, care must be taken when evaluating ${\cal F}^{(d)}(\gamma)$
at the branch point.  The contour chosen for the evaluation of 
Eq. (\ref{trace}) is shown in Fig. \ref{contour}b, where a cut has
been introduced along the negative real axis, and the large arc does not
pass through any of the poles of the integrand.  Following the method of 
Sondheimer and Wilson 
,\cite{sondheimer} we find that the integral in Eq. (\ref{trace})
can be written as a Riemann zeta function plus the sum of the residues at the
poles $s_k = \pm k\pi i$

\begin{eqnarray}
{\cal F}^{(d)}(\gamma) \equiv \int_{\sigma} 
\frac{ds}{2\pi i} \frac{e^{2\gamma s}}{s^d {\rm sinh}(s)} +
\sum_{k} {\rm Res}\left( \frac{e^{2\gamma s_k}}{s_k^d {\rm sinh}(s_k)}\right)~,
\label{fd-half}
\end{eqnarray}
where $\sigma$ denotes the contour FEDC in Fig. \ref{contour}b.  The first
integral in Eq. ({\ref{fd-half}) can be separated into two parts by using an
asymptotic expansion of the integrand at the branch point

\begin{eqnarray}
&&\int_{\sigma}\frac{ds}{2\pi i} \frac{e^{2\gamma s}}{s^d {\rm sinh}(s)} =
\int_{\sigma} \frac{ds}{2\pi i}e^{2\gamma s}\sum_{k=0}^{d+\frac{1}{2}}
\frac{B_{k}}{s^{d+1-k}}
\nonumber \\
&& -
\int_{\sigma} \frac{ds}{2\pi i}e^{2\gamma s}
\left( \sum_{k=0}^{d+\frac{1}{2}} 
\frac{B_k}{s^{d+1-k}} 
-\frac{1}{s^d{\rm sinh}(s)}\right)~.
\end{eqnarray}
where

\begin{eqnarray}
B_k = \frac{1}{k!}\frac{\partial^k}{\partial s^k}
\left(\frac{s}{{\rm sinh}(s)}\right)_{s=0}~.
\end{eqnarray}
The first integral is evaluated by means of Hankel's formula for the gamma
function, viz.,\cite{gradshteyn}

\begin{equation}
\int_{\sigma}\frac{ds}{2\pi i}e^{2\gamma s} s^{-z} =
\frac{(2\gamma)^{z-1}}{\Gamma(z)}~,
\label{gamma}
\end{equation}
and the second integral can be cast as real since the part contributed by the
circle around the origin (see Fig. \ref{contour}b) is zero.  The sum of the
residues is readily computed, and we finally have

\begin{eqnarray}
{\cal F}^{(d)}(\gamma) &=& \sum_{k=0}^{d+\frac{1}{2}} 
\frac{B_k(2\gamma)^{d-k}}{\Gamma(d+1-k)}
+ \frac{1}{\pi}{\cal G}^{(d)}(\gamma)
\nonumber \\
&&(-1)^{d+\frac{1}{2}}\sum_{k=1}^{\infty} \frac{2(-1)^k}{(\pi k)^d}
{\rm cos}(2\pi \gamma k + \pi d/2)
\label{half-fd2}
\end{eqnarray}
where

\begin{eqnarray}
{\cal G}^{(d)}(\gamma) &=&
\int_{0}^{\infty} dy e^{2\gamma y}\left( 
\sum_{k=0}^{d+\frac{1}{2}}
\frac{B_k}{y^{d+1-k}}
- \frac{1}{y^d{\rm sinh}(y)}\right)~.
\label{gd}
\end{eqnarray}
The first term in Eq. (\ref{half-fd2}) gives the TF expression 
for the number density in a homogeneous magnetic field.
The second and third terms are purely quantum-mechanical, and are the origin
of the de Haas-van Alphen effect.\cite{haas}  These oscillatory terms are clearly a
result of the exact treatment of the canonical density matrix, and are not
obtained within the usual TF approximation.\cite{brack} 

Integral values of $d$
allow for an exact treatment of Eq. (\ref{trace}).  Using the contour
shown in Fig. \ref{contour}a, ${\cal F}^{(d)}(\gamma)$ can be expressed in the 
form

\begin{eqnarray}
{\cal F}^{(d)} (\gamma)&=& 
\sum_k{\rm Res}\left(\frac{e^{2\gamma s_k}}
{s_k^{d}{\rm sinh}(s_k)} \right)~,
\label{fd}
\end{eqnarray}
where $s_k = k\pi i~~~~(k=0,\pm 1,\pm 2,...)$.  The
$k=0$ contribution represents the usual TF approximation, 
whereas the terms associated with the poles at $\pm k\pi i$ are 
responsible for the quantum-mechanical de Haas-van Alphen effect.
As in the half-integral case, ${\cal F}^{(d)}(\gamma)$ for even-N will be 
the sum of steady terms ($s_k=0$) plus oscillatory terms ($s_k = \pm k\pi i$):  

\begin{eqnarray}
{\cal F}^{(d)} &=&  A^{(d)}_{k=0} + \nonumber \\ && (-1)^d \sum_{k=1}^{\infty}
\frac{2(-1)^{k}}{(\pi k)^d}{\rm cos}(2\pi\gamma k + \pi d/2)
\label{special_fd}
\end{eqnarray}
where the $A^{(d)}_{k=0}$ terms are given by

\begin{eqnarray}
A^{(d)}_{k=0} &=& \frac{1}{d!}\frac{\partial^d}{\partial s^d} 
\left(\frac{e^{2\gamma s}s}{{\rm sinh}(s)}\right)_{s=0}~.
\end{eqnarray}

In the limit $\omega \rightarrow 0$, the asymptotic
behaviour of ${\cal F}^{(d)}(\gamma)$ for arbitrary $d$ is given by
\cite{note2}

\begin{equation}
{\cal F}^{(d)}(\gamma) = \frac{(2\gamma)^d}{\Gamma(d+1)},
\label{zero-fd}
\end{equation}
and Eq. (\ref{rhod}) reduces to the N-dimensional TF 
theory, viz.,\cite{brack,kventsel}

\begin{equation}
\rho^{(d)}_{\rm TF} = \left(\frac{m}{2\pi\hbar^2}\right)^{d}
\frac{2}{\Gamma(d+1)}\varepsilon_f^d~.
\label{zero-rhod}
\end{equation}

For large magnetic fields, the general behaviour of 
${\cal F}^{(d)}(\gamma)$ for half-integral $d$ is difficult to obtain using
the present treatment. 
Once again, the cause of
this difficulty lies in the branch point at $s=0$ of the complex $s$-plane.
We can circumvent this problem by writing Eq. (\ref{trace}) in the following
form\cite{shoenberg}

\begin{eqnarray}
{\cal F}^{(d)}(\gamma) = 2\sum_{k=0}^{\infty}\int_{\epsilon -i\infty}
^{\epsilon +i\infty} \frac{ds}{2\pi i} \frac{e^{(2\gamma s - 2 k - 1)s}}
{s^d}~,
\label{fd-series}
\end{eqnarray}
where we have used the series expansion

\begin{eqnarray}
\frac{1}{{\rm sinh}(s)} = 2\sum_{k=0}^{\infty} e^{-(2 k + 1)s}~,
\end{eqnarray}
for the sinh(s) function.  Only those terms satisfying $2 k + 1 < 2\gamma $ 
survive the integration in Eq. (\ref{fd-series}), and since 
${\rm max}[2\gamma] = 2$, we have

\begin{eqnarray}
{\cal F}^{(d)}(\gamma) = \frac{2}{\Gamma(d)}(2\gamma -1)^{d-1}~.
\label{fd-high}
\end{eqnarray} 
Equations (\ref{rhod}) and (\ref{fd-high}) provide the high-magnetic
field electronic number density for arbitrary values of $d$.

\section{ENERGY-DENSITY FUNCTIONALS}
\label{functionals}

The kinetic and exchange energy functionals can both be calculated from
the single-particle density matrix, given by Eq. (\ref{den-matrix}).  However,
in the present context, a simplification for the evaluation of the kinetic
energy density is possible.
Our first step is to notice that the Heavyside function may be
represented as the inverse Laplace transform of the exponential function:
\cite{yonei}

\begin{equation}
\theta (\varepsilon_f - \hat{H}) = 
\int_{\epsilon - i\infty}^{\epsilon + i\infty} \frac{d\beta}{2\pi i}
\frac{e^{\beta(\varepsilon_f-\hat{H})}}{\beta}~.
\end{equation}
The kinetic energy density $\varepsilon_{\rm kin}$ may then be expressed
as

\begin{eqnarray}
\varepsilon_{\rm kin} &=& 
\langle \bfr\mid \hat{H}\theta (\varepsilon_f - \hat{H})\mid\bfr \rangle
\nonumber \\
&=& 
\int_{\epsilon - i\infty}^{\epsilon + i\infty} \frac{d\beta}{2\pi i}
\frac{
\langle\bfr\mid \hat{H} e^{\beta(\varepsilon_f-\hat{H})}\mid\bfr\rangle
}{\beta}\nonumber \\
&=&
-\int_{\epsilon - i\infty}^{\epsilon + i\infty} \frac{d\beta}{2\pi i}
\frac{e^{\beta\varepsilon_f}}{\beta}\frac{\partial C(\bfr,\bfr;\beta)}
{\partial \beta}~.
\end{eqnarray}
For arbitrary dimension, $\varepsilon_{\rm kin}$ takes the form

\begin{eqnarray}
\varepsilon^{(d)}_{\rm kin} &=& \left(\frac{m}{4\pi}\right)^{d}\hbar^{d-1}\omega^{d+1}
\int_{\epsilon - i\infty}^{\epsilon + i\infty}
\frac{ds}{2\pi i} e^{2\gamma s} \times \nonumber \\ 
&&\left[
\frac{(d-1)}{s^{d+1}{\rm sinh}(s)} +
\frac{{\rm cosh}(s)}{s^d{\rm sinh}(s)^2} \right] \nonumber \\
&=&
\left(\frac{m}{4\pi}\right)^{d}\hbar^{1-d}\omega^{d+1}
\left[ 2\gamma {\cal F}^{(d)}(\gamma) - {\cal F}^{(d+1)}(\gamma) \right]
\nonumber \\
&=&
\left(\frac{4\pi\hbar^2}{m}\right)\left(
\frac{\rho^{(d)}}{2}\right)^{\frac{d+1}{d}}H^{(d)}(\gamma)~,
\label{kin-functional}
\end{eqnarray}
where 
\begin{equation}
H^{(d)}(\gamma) =
\frac{2\gamma}{\left[{\cal F}^{(d)}\right]^{1/d}} -
\frac{[{\cal F}^{(d+1)}]}{\left[{\cal F}^{(d)}\right]^{(d+1)/d}}~.
\label{hd}
\end{equation}
Equation (\ref{kin-functional}) is valid for arbitrary magnetic field strength.

In the limit of zero
magnetic field, the asymptotic behaviour of $H^{(d)}(\gamma)$ for 
arbitrary $d$ can be deduced from Eq. (\ref{zero-fd}), viz.,

\begin{eqnarray}
H^{(d)}(\gamma) = \frac{\Gamma(d+2)[\Gamma(d+1)]^{\frac{1}{d}} -
[\Gamma(d+1)]^{\frac{d+1}{d}}}{\Gamma(d+2)}~.
\label{hd-low}
\end{eqnarray} 
Thus, using the above value of $H^{(d)}(\gamma)$ in Eq. (\ref{kin-functional}),
we recover the kinetic energy density of the
N-dimensional TF theory. \cite{brack,kventsel}
As the dimension of the space goes to infinity, we obtain

\begin{eqnarray}
H^{(d)}(\gamma) = 1~,
\end{eqnarray}
whereby the kinetic energy density becomes

\begin{eqnarray}
\varepsilon^{(d)}_{\rm kin} &=& 4\pi\left(\frac{\hbar^2}{2m}\right) 
\rho^{(d)}~.
\end{eqnarray}

Substituting Eq. (\ref{fd-high}) into Eq. (\ref{hd}) gives
the high-field limit of $H^{(d)}(\gamma)$, viz.,

\begin{eqnarray}
H^{(d)}(\gamma) &=& 
\frac{4\gamma\Gamma(d)^{\frac{1}{d}}\Gamma(d+1)(2\gamma-1)^{d-1}}
{\Gamma(d+1)2^{\frac{d+1}{d}}(2\gamma -1)^{\frac{d^2-1}{d}}} \nonumber \\
&-&
\frac{2(2\gamma-1)^d\Gamma(d)^{\frac{d+1}{d}}}
{\Gamma(d+1)2^{\frac{d+1}{d}}(2\gamma -1)^{\frac{d^2-1}{d}}}
\label{hd-high}
\end{eqnarray}
In the limit $d \rightarrow \infty$, we obtain

\begin{eqnarray}
H^{(d)}(\gamma) &=& 2\gamma~,
\end{eqnarray}
and the kinetic energy behaves as

\begin{eqnarray}
\varepsilon^{(d)}_{\rm kin} &=& 8\pi\gamma \left(\frac{\hbar^2}{2m}\right)
\rho^{(d)} \nonumber \\
&\simeq&
8\pi\left(\frac{\hbar^2}{2m}\right)
\rho^{(d)}~.
\end{eqnarray}

The other density functional of interest is the
N-dimensional exchange energy $\varepsilon_{\rm ex}$. 
According to the usual prescription, $\varepsilon_{\rm ex}$ is given by
\cite{dreizler}

\begin{equation}
\varepsilon_{\rm ex} = -\frac{e^2}{4} \int d\bfr'\frac{\mid \rho(\bfr,\bfr')\mid^2}
{\mid \bfr - \bfr'\mid}~,
\label{exch-functional}
\end{equation}
where the integration is understood to be over the N-dimensional volume.
The exchange energy density has a much richer $s$-plane topology
than the kinetic energy density because of the essential singularities 
arising from the
${\rm coth}(s)$ term in Eq. (\ref{canonical}).
These essential singularities make it difficult (if not impossible) to
obtain a useful, closed analytical form for the N-dimensional
exchange energy for {\em arbitrary} magnetic fields.
Nevertheless, some analytical progress can be made if we make use of the
following expansion:\cite{gradshteyn}

\begin{eqnarray}
\frac{{\rm exp}(-\alpha{\rm coth}(s))}{{\rm sinh}(s)} =
2 \sum_{n=0}^{\infty} {\rm L}_{n}(2\alpha)e^{-\alpha}e^{-(2n+1)s}~,
\end{eqnarray}
where ${\rm L}_{n}(\alpha)\equiv {\rm L}^{(0)}_{n}(\alpha)$ 
is the Laguerre polynomial.
It is then easy to show that the N-dimensional single-particle density matrix
has the form

\begin{eqnarray}
\rho(\bfr,\bfr') &=& 4\left(\frac{m\omega}{4\pi\hbar}\right)^{d}
e^{-\frac{m\omega}{2\hbar}(x'y-y'x)}\sum_{n=0}^{\infty}{\rm L}_{n}(2\alpha)
e^{-\alpha}\nonumber \\
&&\int_{\varepsilon-i\infty}^{\varepsilon+i\infty}\frac{ds}{2\pi i}
\frac{e^{(2\gamma - 2n - 1)s}e^{-\delta/s}}{s^d}~,
\label{N-single-particle}
\end{eqnarray}
where we have defined

\begin{eqnarray}
\alpha = \frac{m\omega}{4\hbar}\mid\bfr-\bfr'\mid^2~~~;~~~
\delta = \frac{m\omega}{4\hbar}\sum_{i=1}^{2(d-1)}(z_i-z_i')^2~.
\end{eqnarray}
Equation (\ref{N-single-particle}) in Eq. (\ref{exch-functional}) yields,
in principle, the N-dimensional exchange energy for arbitrary magnetic field.
Of course, the comments made in Sec. \ref{density-matrix} with regard
to the electron number density also apply to Eq. (\ref{N-single-particle}).

Below, we illustrate the utility of our formalism through specific examples
in the physically relevant 3D and 2D systems.

\subsection{N=3}
\label{3d}

The electron number density in three-dimensions can be obtained from
Eqs. (\ref{rhod}) and (\ref{half-fd2}):

\begin{eqnarray}
&&\rho^{(3/2)}(\gamma)= \frac{1}{4} \left(\frac{m\omega}{\pi\hbar}\right)^{3/2}
\left[\frac{4}{3\sqrt{\pi}}(2\gamma)^{3/2} - \frac{1}{6\sqrt{2\pi \gamma}}
\right. \nonumber \\ 
&&+\left. \frac{1}{\pi}{\cal G}^{(3/2)}(\gamma)+
\sum_{k=1}^{\infty} \frac{2}{\pi^{3/2}}\frac{(-1)^{k}}{k^{3/2}}
{\rm cos}\left(2k\pi \gamma + \frac{3\pi}{4}\right)
\right]~,
\label{density3d}
\end{eqnarray}
where ${\cal G}^{(3/2)}(\gamma)$ is given by Eq. (\ref{gd}).
In the zero field limit, we can make use of Eq. (\ref{zero-rhod}) to get

\begin{eqnarray}
\rho^{(3/2)}(\gamma) = \frac{1}{3\pi^2}
\left(\frac{2m}{\hbar^2}\right)^{3/2} \varepsilon_f^{3/2}~,
\label{rho3dlow}
\end{eqnarray}
and in the high-field limit, Eqs. (\ref{rhod}) and (\ref{fd-high}) yield, 

\begin{eqnarray}
\rho^{(3/2)}(\gamma) = \left(\frac{2m^{3}\omega^2}{\pi^4\hbar^4}
\right)^{1/2}(\varepsilon_f - \hbar\omega/2)^{1/2}~.
\label{rho3dhigh}
\end{eqnarray}
Both of these limiting cases are in full agreement with the results expected
for a 3D homogeneous electron gas.\cite{dreizler}

The kinetic energy density is most easily obtained by using the second form
of Eq. (\ref{kin-functional}) with $d=3/2$, viz.,

\begin{eqnarray}
\varepsilon_{\rm kin} &= &
\left(\frac{m}{4\pi}\right)^{3/2}\hbar^{-1/2}\omega^{5/2}
\left[ 2\gamma {\cal F}^{(3/2)} - {\cal F}^{(5/2)} \right]~.
\end{eqnarray}
The functions ${\cal F}^{(3/2)}(\gamma)$ and ${\cal F}^{(5/2)}(\gamma)$
are readily obtained from Eqs. (\ref{half-fd2}), and (\ref{gd}), and 
therefore

\begin{eqnarray}
\varepsilon_{\rm kin} &=& 
\left(\frac{m}{4\pi}\right)^{3/2}\hbar^{-1/2}\omega^{5/2}
\left[
\frac{4(2\gamma)^{5/2}}{5\sqrt{\pi}} 
+ \frac{(2\gamma)^{1/2}}{6\sqrt{\pi}} \right. \nonumber \\
&+& \left.  \frac{2\gamma}{\pi} {\cal G}^{(3/2)}(\gamma) - 
\frac{1}{\pi}{\cal G}^{(5/2)}(\gamma) \right.
\nonumber \\
&+& \left.
\sum_{k=1}^{\infty} \frac{4\gamma(-1)^{(k+1)}}{(\pi k)^{3/2}}
{\rm sin}(2k\pi \gamma + \frac{\pi}{4}) \right. \nonumber \\
&+&\left. \left.
\sum_{k=1}^{\infty} \frac{2(-1)^k}{(\pi k)^{5/2}}{\rm sin}(2k\pi \gamma
- \frac{\pi}{4}) \right. \right ]~.
\label{kinetic3d}
\end{eqnarray}
The last four terms in Eq. (\ref{kinetic3d}) are again quantum effects and
are the origin of the de Haas-van Alphen effect.  

Using Eq. (\ref{hd-low}) in Eq. (\ref{kin-functional}) yields the kinetic 
energy density in the zero-field limit

\begin{eqnarray}
\varepsilon_{\rm kin} = \frac{3}{5}(3\pi^2)^{2/3}
\left(\frac{\hbar^2}{2m}
\right)\rho^{5/3}~,
\label{3dkinlow}
\end{eqnarray}
and in the high-field regime, Eq. (\ref{hd-high}) in Eq. (\ref{kin-functional})
gives

\begin{eqnarray}
\varepsilon_{\rm kin} = \frac{\hbar\omega}{2}\rho +
\frac{2\pi^4\hbar^4}{3m^3\omega^2}\rho^3~.
\label{3dkinhigh}
\end{eqnarray}
Equation (\ref{3dkinlow}) is the well known TF result for a degenerate
noninteracting 3D electron gas,\cite{dreizler} while Eq. (\ref{3dkinhigh}) agrees with
the result of Yonei {\em et al.,}\cite{yonei} and
Banerjee {\em et al.}\cite{banerjee}  Note that the 
functional form of the kinetic energy density is very different 
in the low and high-magnetic field limits.  
In particular, the $\rho^3$ term 
appearing in Eq. (\ref{3dkinhigh}) is a 
reflection of the one dimensional nature of the system along the field
direction.

In order to evaluate the exchange energy density,
we require explicit expressions for the single-particle density matrix.
In principle, Eq. (\ref{N-single-particle}) provides such an expression. 
However, it is rather difficult to obtain the exchange energy for 
arbitrary magnetic field in 3D,\cite{danz} so we focus here only on the
the zero and high-field limits.

In the zero-field limit, we set ${\rm sinh}(s) \approx s$, and
${\rm coth}(s) \approx s^{-1}$.  Substituting these expressions into
Eq. (\ref{canonical}), and performing the inverse Laplace transform, we obtain

\begin{eqnarray}
\rho(\bfr,\bfr') &=& \frac{\sqrt{\varepsilon_f}}{a\sqrt{\pi}}
{\rm cosh}(2\sqrt{a\varepsilon_f}) \nonumber \\
&-& \frac{1}{2\sqrt{a^3\pi}} {\rm sinh}(2\sqrt{a\varepsilon_f})~, 
\end{eqnarray}
where $a \equiv (m/2\pi\hbar^2)\mid \bfr - \bfr'\mid^2$.  
This expression leads back to Eq. (\ref{rho3dlow}) upon taking the limit 
$\bfr \rightarrow \bfr'$.  Equation
(\ref{exch-functional}) can now be used to yield the exchange energy
functional.  This is a straightforward, but tedious calculation, and since 
it only leads back to the well know expression for a 3D free-electron gas, 
we simply quote the result here\cite{dreizler}

\begin{eqnarray}
\varepsilon_{\rm ex} = -\frac{3}{4}\left(\frac{3}{\pi}\right)^{1/3} e^2
\rho^{4/3}~.
\label{exch3dlow}
\end{eqnarray}
The high-field limit for the single-particle density matrix is obtained by
replacing ${\rm sinh}(s) \approx 1/2{\rm exp}(-s)$, and
${\rm coth}(s) \approx 1$ in Eq. (\ref{canonical}).  After performing the
inverse Laplace transform of the canonical density matrix, we obtain

\begin{eqnarray}
\rho(\bfr,\bfr') &=& \frac{2\hbar\omega}{\sqrt{2}\pi^2}
\left(\frac{m}{2\pi\hbar^2}\right)^{3/2}\frac{1}{\mid z-z'\mid}\nonumber
\\ &\times&
{\rm sin}(\sqrt{2(\varepsilon_f-\hbar\omega/2)}\mid z-z'\mid) \nonumber \\
&\times&
{\rm exp}\left(-i\frac{m\omega}{2\hbar}(x'y-y'x)\right)\nonumber \\
&\times&
{\rm exp}\left(-\frac{m\omega}{4\hbar}((x-x')^2+(y-y')^2)\right)~.
\label{rhorrp3dhigh}
\end{eqnarray}
(The high-field limits of the number density can be obtained
from Eq. (\ref{rhorrp3dhigh}) by taking the limit $\bfr \rightarrow \bfr'$).
The exchange energy density can now be calculated by substituting Eq.
(\ref{rhorrp3dhigh}) into Eq. (\ref{exch-functional}) and transforming
over to cylindrical coordinates: 

\begin{eqnarray}
\varepsilon_{\rm ex} &=& -\frac{4e^2}{(2\pi)^3}
\left( \frac{m\omega}{\hbar}\right)^2\int_{0}^{\infty} dz z^{-2} 
{\rm sin}^2\left[\left(\frac{4\pi^4\hbar^4\rho^2}{m^3\omega^2}\right)^{1/2}
z\right]\nonumber \\
&\times&\int_{0}^{\infty}dr~ r(r^2+z^2)^{-1/2}{\rm exp}
\left(-\frac{m\omega}{2\hbar}\right)r^2~.
\end{eqnarray}
Fortunately, Danz and Glasser\cite{danz} have already considered a 
similar calculation,
and after several manipulations in accordance with their work, we obtain

\begin{eqnarray}
\varepsilon_{\rm ex} &=&
-\frac{e^2}{4\pi^3}\left(\frac{m\omega}{\hbar}\right)^2\left[
{\rm p} \int_{0}^{\infty} dx~x{\rm ln}(x^2){\rm exp}(-{\rm p}x^2)\right.
\nonumber \\
&-& \left.
{\rm p}\int_{0}^{\infty} dx~x {\rm ln}(x^2+1){\rm exp}(-{\rm p}x^2)\right.
\nonumber \\
&+& \left.
2{\rm p} \int_{0}^{\infty} dx~{\rm exp}(-{\rm p}x^2) {\rm tan}^{-1}(x^{-1})
\right]~,
\label{exch3dhigh}
\end{eqnarray}
where 
\begin{eqnarray}
{\rm p} = \frac{8\pi^4\hbar^3}{m^3\omega^3}\rho^2~.
\end{eqnarray}
The asymptotic behaviour of Eq. (\ref{exch3dhigh}) in the low-density
limit is given by

\begin{eqnarray}
\varepsilon_{\rm ex} &\approx& \left(\frac{e^2}{8\pi^3}\right)\left(
\frac{m\omega}{\hbar}\right)^2{\rm p}\left[{\rm ln}({\rm p}) - (3-C)\right]~,
\end{eqnarray}
where $C$ is Euler's constant.  In the high-density regime
(where the semiclassical TF theory is applicable), Eq. (\ref{exch3dhigh})
has the form

\begin{eqnarray}
\varepsilon_{\rm ex} &\approx& -\left(\frac{e^2}{8\pi^3}\right)\left(
\frac{m\omega}{\hbar}\right)^2 {\rm p}^{1/2}~.
\end{eqnarray}
Similar to the kinetic energy density functional, the behaviour
of the exchange energy density is markedly different in the low and high
magnetic field limits.  


\subsection{N=2}
\label{2d}
Since $d$ is integral, we can make use of Eq. (\ref{special_fd})
to obtain the electron number density in terms of elementary functions, viz.,

\begin{eqnarray}
\rho^{(1)}(\gamma) &=& \left(\frac{m\omega}{2\pi\hbar}\right)
{\cal F}^{(1)}(\gamma)\nonumber \\
&=&
\left(\frac{m\omega}{2\pi\hbar}\right)\left[
2\gamma + \sum_{k=1}^{\infty} \frac{2(-1)^k}{\pi k}{\rm sin}(2\pi k \gamma)
\right]
~,
\end{eqnarray}
The low and high-field limits of the number density are given by

\begin{eqnarray}
\rho^{(1)}(\gamma) = \frac{m\varepsilon_f}{\pi\hbar^2}~,
\label{rho2dlow}
\end{eqnarray}
and
\begin{eqnarray}
\rho^{(1)}(\gamma) = \frac{m\omega}{\pi\hbar}~,
\label{rho2dhigh}
\end{eqnarray}
respectively.  

The kinetic energy density can also be expressed in terms of elementary
functions, viz.,

\begin{eqnarray}
\varepsilon_{\rm kin} &=& \left(\frac{m\omega^2}{4\pi}\right)
\left[2\gamma {\cal F}^{(1)}(\gamma) - {\cal F}^{(2)}(\gamma)\right]
\nonumber \\
&=& \left(\frac{m\omega^2}{4\pi}\right)\left[
2\gamma^2 + \frac{1}{6} + 
\sum_{k=1}^{\infty} \frac{4\gamma (-1)^k}{\pi k}{\rm sin}(2\pi k\gamma)
\right. \nonumber \\
&+& \left.
\sum_{k=1}^{\infty} \frac{2(-1)^k}{(\pi k)^2}{\rm cos}(2\pi k\gamma)
\right]\nonumber \\
&=&
\left(\frac{\hbar^2}{m}\right)\pi \left[\rho^{(1)}(\gamma)\right]^2
H^{(1)}(\gamma)\nonumber \\
&=&
\left(\frac{\hbar^2}{2m}\right)\pi \left[\rho^{(1)}(\gamma)\right]^2~,
\label{kin2d}
\end{eqnarray}
where we have used the fact that
$H^{(1)}(\gamma) = 1/2$ $\forall \gamma$.  Therefore, 
in remarkable contrast to the 3D case, the
the 2D kinetic energy density has {\em exactly} the same
functional form regardless of the magnetic field strength; all of the
de Haas-van Alphen type behaviour is buried in the $\rho^{(1)}(\gamma)$ 
dependence.
This result is peculiar to 2D, since in all other dimensions,
$H^{(d)}(\gamma)$ is generally a function of $\omega$.
For example, in four-dimensions, we have from Eq. (\ref{hd-low})

\begin{eqnarray}
\varepsilon_{\rm kin} = 
\pi\left(\frac{\hbar^2}{2m}\right)\frac{8}{3}\rho^{3/2}~,
\end{eqnarray}
in the zero-field limit, and from Eq. (\ref{hd-high})
\begin{eqnarray}
\varepsilon_{\rm kin} =
\frac{\pi^2\hbar^3}{m^2\omega}\rho^2 + \frac{\hbar\omega}{2}\rho~,
\end{eqnarray}
in the high-field limit.

The exchange energy is calculated along the same lines as Sec. \ref{3d}.
The single-particle density matrix for arbitrary magnetic field is obtained
by setting $d=1$ in Eq. (\ref{N-single-particle}), and is given by

\begin{eqnarray}
\rho(\bfr,\bfr') &=& \frac{m\omega}{\pi\hbar}
{\rm exp}\left(\frac{-im\omega}{2\hbar}(x'y-y'x)\right)\times\nonumber \\
&&{\rm exp}\left(-\frac{m\omega}{4\hbar}
\mid\bfr-\bfr'\mid^2\right)\times
\nonumber \\ 
&&{\rm L}^{(1)}_{n_f}\left(
\frac{m\omega}{2\hbar}\mid\bfr-\bfr'\mid^2\right)~,
\label{exch2d-exact}
\end{eqnarray}
where we have summed only up to the Fermi level $n_f$, and used the 
finite summation relation \cite{gradshteyn}

\begin{eqnarray}
\sum_{n=0}^{m} {\rm L}^{(k)}_n(x) = {\rm L}^{(k+1)}_{m}(x)~.
\end{eqnarray}
The exchange energy in 2D can be evaluated from Eq. (\ref{exch-functional}),
provided that we make use of the identity\cite{gradshteyn}

\begin{eqnarray}
\left[{\rm L}^{(1)}_n(x)\right]^2 &=&
\frac{\Gamma(2+n)}{n!}\sum_{k=0}^{\infty} 
\frac{(2n - 2k)!(2k)!}{\Gamma(2-k)} 
\frac{{\rm L}^{(2)}_{2k}(2x)}{(n-k)!}~.
\end{eqnarray}
Upon substituting Eq. (\ref{exch2d-exact}) into Eq. (\ref{exch-functional}),
we obtain after some tedious algebra

\begin{eqnarray}
\varepsilon_{\rm ex} &=& -\frac{e^2}{\pi}
\left(\frac{m\omega}{2\hbar}\right)^{3/2}n_f\sum_{k=0}^{n_f}\nonumber \\
&&\frac{
\Gamma(n_f + 2 + k)\Gamma(2k+1/2)}
{\Gamma(n_f+1-k)\Gamma(2k+2)\Gamma(k+2)\Gamma(k+1)}\times \nonumber \\
&&F[-(n_f-k),(2k+1/2);(2k+2);2]~,
\label{exch2d}
\end{eqnarray}
where $F(a,b;c;d)$ denotes the hypergeometric series.\cite{gradshteyn}
The limiting forms of Eq. (\ref{exch2d}) in the zero and high-field limits
can be made more transparent by first examining at the limiting forms of
$\rho(\bfr,\bfr')$ in Eq. (\ref{den-matrix}).

In the zero-field limit, we have

\begin{eqnarray}
\rho(\bfr,\bfr') &=& \left(\frac{1}{\pi}\right) \frac{1}{\mid\bfr -\bfr'\mid}
k_f J_1(k_f \mid \bfr - \bfr'\mid)~,
\label{rhorrp2dlow}
\end{eqnarray}
where the Fermi wavevector $k_f = (2m\varepsilon_f/\hbar^2)^{1/2}$.
Once again, Eq. (\ref{rho2dlow}) is recovered by taking
the limit $\bfr \rightarrow \bfr'$ in Eq. (\ref{rhorrp2dlow}).
The exchange energy in this limit is given by an application of Eq.
(\ref{exch-functional}), viz.,

\begin{eqnarray}
\varepsilon_{\rm ex} &=& -\frac{2}{3}e^2\frac{1}{\pi^2}k_f^3 \nonumber \\
&=& -\frac{4}{3}\sqrt{\frac{2}{\pi}}e^2\rho^{3/2}~.
\label{exch2dlow}
\end{eqnarray}
This is the usual 2D free-electron gas result.  In the high field limit,
the single-particle density matrix takes the form

\begin{eqnarray}
\rho(\bfr,\bfr') &=& \frac{m\omega}{\pi\hbar}
{\rm exp}\left(-i\frac{m\omega}{2\pi\hbar}(x'y-y'x)\right) \nonumber \\
&\times&
{\rm exp}\left(-\frac{m\omega}{4\pi\hbar}((x-x')^2+(y-y')^2)\right)~,
\label{exch2dhigh}
\end{eqnarray}
and upon using Eq. (\ref{exch-functional}), we find the simple result

\begin{eqnarray}
\varepsilon_{\rm ex} &=& -\frac{\pi e^2}{2\sqrt{2}}\rho^{3/2}~.
\end{eqnarray}
Equations (\ref{exch2dlow}), and (\ref{exch2dhigh}) should be contrasted
with Eqs. (\ref{exch3dlow}), and (\ref{exch3dhigh}).  Obviously, the 
key difference between the two sets of functionals is the universal form
assumed by the exchange energy in 2D.  Specifically, the exchange energy
functional in 2D is explicitly independent of $\gamma$
in limit of strong magnetic fields.  

\section{CURRENT-DENSITY FUNCTIONAL THEORY IN STRONG MAGNETIC FIELDS:\\
THE TFD APPROXIMATION}
\label{cdft}

In this section, we take a pedestrian approach to the construction
of a current density functional theory applicable to
inhomogeneous electronic systems in 
strong magnetic fields.   The foundations of the CDFT formalism have already
been provided by VR \cite{vignale}, 
and we invite the interested reader to refer to this work
for rigorous proofs and details.

Consider an {\em inhomogeneous} electronic system in a magnetic field.  
Within the TFD approximation, we assume that the non-uniform electronic
system can be locally approximated as a uniform electron gas characterized
by the density $\rho(\bfr)$, local Fermi wavevector $k_f(\bfr)$, and
canonical (or ``conjugate'') current density ${\bf j}_{p}(\bfr)$ (see below).
Thus, a local approximation to the energy functional is obtained by summing
over all N-dimensional volume elements, treating each N-dimensional
volume element as independent;
this is the local-density
approximation (LDA) in which the density is assumed to vary slowly on the
scale of the magnetic length $\ell_{B} = (\hbar c/eB)^{1/2}$.

The local approximation to the total ground state energy functional for 
an N-dimensional many-electron system
characterized by a scalar potential $v(\bfr)$ and vector potential
${\bf A}(\bfr)$ is given by

\begin{eqnarray}
E_{{\rm TFD}} &=& \int d\bfr~\varepsilon[\rho(\bfr),\mid \bbox{\nu} (\bfr)\mid]
+ \frac{m}{2}\int d\bfr~\frac{\mid {\bf j}(\bfr)\mid^2}{\rho(\bfr)}
\nonumber \\
&+& E_{\rm H}[\rho(\bfr)] + \int d\bfr~\rho(\bfr) v(\bfr)~,
\label{cdft-functional}
\end{eqnarray}
where $E_{\rm H}$ is the nonlocal Hartree contribution, and 
$\varepsilon = \varepsilon_{\rm kin} + \varepsilon_{\rm ex}$
is the total (gauge-invariant) energy density of the uniform electron 
gas in a fictitious magnetic field, defined by the vorticity 
$\bbox{\nu} (\bfr)$ 

\begin{eqnarray}
\bbox{\nu} (\bfr) = \nabla \times \frac{{\bf j}_{p}(\bfr)}{\rho(\bfr)}~.
\label{vorticity}
\end{eqnarray}
The physical orbital current density is given by

\begin{eqnarray}
{\bf j}(\bfr) = {\bf j}_{p}(\bfr) + \frac{e}{mc}\rho(\bfr){\bf A}(\bfr)~,
\end{eqnarray}
where

\begin{equation}
{\bf j}_{p}(\bfr) \equiv \frac{\hbar}{m}{\rm Im}
[\nabla \rho(\bfr,\bfr')\mid_{r=r'}]~.
\end{equation}
The rationale behind considering the total energy $\varepsilon$ in a fictitious
magnetic field is as follows:  For a given local value of ${\bf j}_{p}$, we
construct a fictitious vector potential ${\bf A}_{\nu}(\bfr)$ which produces
${\bf j}_{p}$ in a uniform electron gas of density $\rho$.  The potential
${\bf A}_{\nu}(\bfr)$ couples only to the current and density.
Since we must have ${\bf j} = 0$ everywhere within a uniform electron gas, 
\cite{landau} we find

\begin{equation}
{\bf A}_{\nu}(\bfr) = -\frac{mc}{e} \frac{{\bf j}_{p}}{\rho(\bfr)}.
\end{equation}
The  magnetic field associated with ${\bf A}_{\nu}(\bfr)$ is nothing more than
the vorticity defined in Eq. (\ref{vorticity}).
The relationship between the fictitious magnetic field and the true
magnetic field is given by

\begin{equation}
\bbox{\nu} (\bfr) = \bbox{\omega} -  \nabla \times 
\frac{{\bf j}(\bfr)}{\rho(\bfr)}.
\label{vorticity2}
\end{equation}
Therefore, the functional $\varepsilon$ is simply constructed by considering the
Euler equation for a homogeneous electron gas in a fictitious magnetic field
$\bbox{\nu}(\bfr)$, viz.,

\begin{eqnarray}
\frac{\partial \varepsilon[\rho(\bfr),\mid\bbox{\nu}\mid]}
{\partial \rho(\bf(r)} = \mu = \mu_{\rm kin} + \mu_{\rm ex}
\label{chemical-potential}
\end{eqnarray}
where the chemical potential $\mu$ now consists of kinetic plus exchange
contributions to the free electron gas in the fictitious magnetic field.

The minimization of Eq. (\ref{cdft-functional}) is done first with respect to 
${\bf j}(\bfr)$ (or equivalently ${\bf j}_{p}$, since $\rho$ and ${\bf A}$
are held constant).  This leads to an Euler-Lagrange equation for the
equilibrium orbital current density, which may be cast in terms of
the density $\rho(\bfr)$.  A substitution of the resulting expression
for the orbital current density into Eq. (\ref{cdft-functional}) yields
a functional, $E_{\rm TFD}$, solely in terms of $\rho(\bfr)$.  A minimization of 
this functional gives the Euler-Lagrange equation for the ground state 
density distribution of the many-electron system.

Below, we treat separately the 3D and 2D systems within the CDFT, 
restricting our attention to the case of strong magnetic fields.  
The energy-density functionals obtained in Secs. \ref{3d} and \ref{2d} 
now find relevance in the construction of the total energy density
$\varepsilon = \varepsilon_{\rm kin} + \varepsilon_{\rm ex}$.

\subsection{N=3}

The total energy density $\varepsilon$ is obtained from 
Eq. (\ref{chemical-potential}), with
\begin{equation}
\mu = \frac{\hbar\mid\bbox{\nu}\mid}{2} + 
\frac{2\hbar^4\pi^4}{m^3\mid\bbox{\nu}\mid^2}\rho^2 -
e^2\left(\frac{m}{8\pi^2\hbar}\right)^{\frac{1}{2}}\mid\bbox{\nu}\mid^{\frac{1}{2}}~,
\label{3dchemical-potential}
\end{equation}
where the last term is the exchange
contribution to the chemical potential.  Using 
Eq. (\ref{3dchemical-potential}), we immediately obtain

\begin{eqnarray}
\varepsilon[\rho,\mid\bbox{\nu}\mid] &=& \frac{\hbar}{2}\mid \bbox{\nu}\mid
\rho + \frac{2\pi^4\hbar^4}{3m^3\mid \bbox{\nu}\mid^2}\rho^3 \nonumber \\
&-& e^2 \left(\frac{m}{8\pi^2\hbar}\right)^{1/2} \mid \bbox{\nu} \mid^{1/2}
\rho~.
\label{total3d}
\end{eqnarray} 
Notice that Eq. (\ref{total3d}) is
formally equivalent to replacing
$\omega \rightarrow \mid \bbox{\nu}\mid$ in Eqs. (\ref{3dkinhigh})
and (\ref{exch3dhigh}) (e.g., the total energy density in the {\em fictitious}
magnetic field).
As discussed above, a variation of the $E_{\rm TFD}$ with respect to 
${\bf j}$ gives

\begin{eqnarray}
{\bf j} &=& \frac{1}{m}\nabla\times
\frac{\partial \varepsilon}{\partial \nu}\nonumber \\
&=& \left(
\frac{\hbar}{2m} - \frac{4\pi^4\hbar^4}{m^4}\frac{\rho^2}{\mid \nu\mid^3}
\right. \nonumber \\
&-& \left. \frac{e^2}{2}\left(\frac{m}{8\pi^2\hbar\mid\nu\mid}\right)^{1/2}
\right) \nabla \rho \times \hat{\bf z}~.
\label{orbital-j}
\end{eqnarray}
In the limit of very strong fields, we obtain

\begin{equation}
{\bf j} = \frac{\hbar}{2m}\nabla\rho\times\hat{\bf z}~,
\label{exact-orbital-j}
\end{equation}
which is exactly the result of Vignale and Skudlarski.\cite{skudlarski}
The vorticity can now be developed by using Eq. (\ref{orbital-j}) in
Eq. (\ref{vorticity2}).  
Inserting the resulting expression for the vorticity into Eq. (\ref{total3d})
gives all of the necessary components for the complete characterization of the
ground state energy functional, Eq. (\ref{cdft-functional}).

One can easily improve on the above TFD theory by including gradient
corrections (the so-called von Weizs\"acker corrections \cite{weizsacker}) 
in the kinetic
energy density.  The explicit inclusion of such gradient terms 
would take into account the
magnetic-induced anisotropy of the density along, and perpendicular to the 
field direction, thereby giving a more realistic treatment of the
system in a strong magnetic field.
Such a calculation has already been performed by
Mazzolo {\it et al.,} \cite{mazzolo} but without the exchange interaction.  
Their final
result for the total energy functional is valid within the Thomas-Fermi-von
Weizsacker approximation, and is given by

\begin{eqnarray}
E_{\rm TFvW} &=& \int d\bfr~\left(
\frac{\hbar\omega}{2}\rho + \frac{2\pi^4\hbar^4}{3m^3\omega^2}\rho^3 
\right. \nonumber \\
&-& \left.
\frac{\hbar^2}{8m}\frac{\mid\nabla_{\perp}\rho\mid^2}{\rho} +
\frac{3\pi^4\hbar^5}{m^4\omega^3}\rho\mid\nabla_{\perp}\rho\mid^2\right.
\nonumber \\
&-& \left.
8m\left(\frac{\pi^4\hbar^4}{m^4\omega^3}\right)^2\rho^3\mid
\nabla_{\perp}\rho\mid^2 - \frac{\hbar^2}{24m}
\frac{\mid\nabla_{\parallel}\rho\mid^2}{\rho}
\right) \nonumber \\
&+& \int d\bfr~\rho v(\bfr)~,
\label{mazzolo-tfdw}
\end{eqnarray}
where $\nabla_{\perp}$ and $\nabla_{\parallel}$ are the gradients perpendicular
and parallel to the magnetic field direction respectively.
Our formalism offers an improvement over
Eq. (\ref{mazzolo-tfdw}) by including the effects of the exchange interaction.

\subsection{N=2}

The determination of the total energy density $\varepsilon$ in 2D
is formally the same as in the 3D case.  However, the simple
substitution $\omega \rightarrow \mid\bbox{\nu}\mid$ in Eqs.
(\ref{kin2d}) and (\ref{exch2dhigh}) for the present scenario
is quite ambiguous.  Obviously, this is because the energy-density
functionals for ${\rm N} = 2$ are explicitly independent of 
the magnetic
field in the strong-field regime.  So, unlike the 3D case, we have to be
a little more careful when constructing the 2D total energy density 
$\varepsilon$.

First, we define the chemical 
potential for the homogeneous 2D electron gas in the fictitious magnetic field:

\begin{eqnarray}
\mu = \frac{\hbar\mid\bbox{\nu}\mid}{2} 
-\frac{\pi e^2}{2\sqrt{2}}\left(\frac{m}{2\pi\hbar}\right)^{1/2}
\mid\bbox{\nu}\mid^{1/2}
\end{eqnarray}
An application of Eq. (\ref{chemical-potential}) yields

\begin{eqnarray}
\varepsilon[\rho,\mid\bbox{\nu}\mid] =
\frac{\hbar\mid\bbox{\nu}\mid}{2}\rho - 
\frac{\pi e^2}{2\sqrt{2}}\left(\frac{m}{2\pi\hbar}\right)^{1/2}
\mid\bbox{\nu}\mid^{1/2}
\rho~.
\label{total2d}
\end{eqnarray}
If we now use Eq. (\ref{rho2dhigh}) with
$\omega \rightarrow \mid\bbox{\nu}\mid$, Eq. (\ref{total2d}) may be cast in
terms of the density $\rho$, viz.,

\begin{eqnarray}
\varepsilon[\rho,\mid\bbox{\nu}\mid] =
\pi\left(\frac{\hbar^2}{2m}\right) \rho^2 - 
\frac{\pi e^2}{2\sqrt{2}}\rho^{3/2}~.
\label{kin-fic2d}
\end{eqnarray}
Recall that in 3D, the magnetic-TFD
theory is readily distinguished from the non-magnetic theory because 
the local 3D energy functionals enjoy very different behaviour in the low and 
high-field limits.  Conversely, the local 2D energy-density 
functionals have no such signature.

The current density is obtained with the aid of Eq. (\ref{total2d}), viz.,

\begin{eqnarray}
{\bf j} &=& \frac{1}{m} \nabla \times \frac{\partial\varepsilon}
{\partial \nu} \nonumber \\
&=& \left( \frac{\hbar}{2m} - \frac{e^2}{8m}\left(
\frac{m}{2\pi\hbar\mid\bbox{\nu}\mid}\right)^{\frac{1}{2}}\right)
\nabla \rho\times \hat{\bf z} \nonumber \\
&\simeq&
\left( \frac{\hbar}{2m} - \frac{e^2}{8m}\left(
\frac{m}{2\pi\hbar\omega}\right)^{\frac{1}{2}}\right)
\nabla \rho\times \hat{\bf z}~.
\label{current-2d}
\end{eqnarray}
In the limit of very strong fields we again recover 
Eq. (\ref{exact-orbital-j}).  MacDonald and Girvin \cite{girvin}
have shown some time ago that
Eq. (\ref{exact-orbital-j}) is a special case of a powerful identity
expressing the single-particle density matrix in terms of the density.
Indeed, for a 2D electron gas in a uniform magnetic field, 
Eq. (\ref{exact-orbital-j}) is the {\em exact} orbital current
density distribution when only the lowest Landau level is occupied.

The recent interest in the optical, and electronic properties of the 2D 
electron gas in a
strong magnetic field (e.g., quantum-hall effect \cite{prange}, 
2D Wigner crystallization \cite{wigner},
edge reconstruction in quantum dots \cite{reimann})
provides adequate motivation
for the explicit calculation of Eq. (\ref{cdft-functional})
in two-dimensions.

The vorticity is determined from Eq. (\ref{current-2d}) and
Eq. (\ref{vorticity}), viz.,

\begin{eqnarray}
\mid\bbox{\nu}(\bfr)\mid &=& \omega - \left(
\frac{\hbar}{2m} - \frac{\pi e^2}{4\sqrt{2}m}
\left(\frac{m}{2\pi\hbar\omega}\right)^{1/2}
\right) \nonumber \\
&\times&
\frac{\mid\nabla_{\perp}\rho\mid^2}{\rho^2}~, 
\end{eqnarray}
where terms of the form $\nabla^2_{\perp}\rho$ are ignored, as they eventually
integrate to zero.
Using this expansion in Eq. (\ref{total2d}), and with 
Eq. (\ref{current-2d}) for the current, we obtain

\begin{eqnarray}
\varepsilon + \frac{m}{2}\frac{\mid{\bf j}\mid^2}{\rho} &=&
\frac{\hbar\omega}{2}\rho - \frac{\pi e^2}{2\sqrt{2}}
\left(\frac{m}{\pi \hbar}\right)^{\frac{1}{2}}\omega^{1/2}\rho\nonumber \\
&-&
\left[\frac{\hbar^2}{8m} - \frac{e^2}{8}
\left(\frac{\pi\hbar}{2m\omega}\right)^{\frac{1}{2}} + \frac{\pi e^4}{64 \hbar\omega}
\right] \nonumber \\
&\times&
\frac{\mid \nabla_{\perp}\rho\mid^2}{\rho}~.
\label{eplusj}
\end{eqnarray}
Using Eq. (\ref{eplusj}) in Eq. (\ref{cdft-functional}), we finally obtain
for the total energy functional, $E_{\rm TFD}$

\begin{eqnarray}
E_{\rm TFD} &=&
\int d\bfr~\left[
\frac{\hbar\omega}{2}\rho - \frac{\pi e^2}{2\sqrt{2}}
\left(\frac{m}{\pi \hbar}\right)^{\frac{1}{2}}\omega^{1/2}\rho \right. \nonumber \\
&-& \left(
\frac{\hbar^2}{8m} - \frac{e^2}{8}
\left(\frac{\pi\hbar}{2m\omega}\right)^{\frac{1}{2}} + \frac{\pi e^4}{64 \hbar\omega}
\right)\frac{\mid \nabla_{\perp}\rho\mid^2}{\rho} \nonumber \\
&+& \left.
v(\bfr)\rho(\bfr) +
\frac{e^2}{2}\int d\bfr'\frac{\rho(\bfr)\rho(\bfr')}{\mid\bfr-\bfr'\mid}
\right]~.
\label{E-tfd}
\end{eqnarray}
Notice that the first two terms have the same form as the
usual zero-field TFD energy functional (the kinetic term is {\em exactly}
the same as the zero-field TF term).  Extending Eq. (\ref{E-tfd}) to
include gradient corrections in the kinetic energy is not possible because
the analogous 2D von Weizs\"acker term vanishes identically.\cite{brack} 
It is interesting to note, however, that the current density naturally 
provides a von Weizs\"acker-like (vW) gradient correction, viz.,

\begin{eqnarray}
E_{\rm vW} =- \frac{\hbar^2}{2m} \int d\bfr~ \frac{\mid\nabla_{\perp}\rho\mid^2}
{4\rho}~,
\end{eqnarray}
with a coefficient that is 1/9-th the magnitude of
the 3D, zero-field TFDvW theory.\cite{brack}  
It is well known that the von Weizs\"acker
term provides the smooth exponential fall-off for the density distribution as
$\bfr \rightarrow \infty$.  

Equation (\ref{E-tfd}) is applicable to any smooth external confining
potential $v(\bfr)$.  For example, choosing
$v(\bfr) = 1/2m\omega_0r^2$ in Eq. (\ref{E-tfd})
would provide a scheme for calculating the ground state density 
distribution of a parabolically confined quantum dot in a strong magnetic 
field.

\section{CONCLUSIONS}
\label{conclusions}

We have presented a general approach for the calculation of the exact 
local-density
functionals for an N-dimensional electronic system in a magnetic field.
We have found that 2D systems enjoy a preferred status over 
other dimensions, in that the form of the 2D kinetic and exchange energy 
functionals are explicitly
independent of the magnetic field.
We have also provided an outline for the construction of a CDFT for
3D and 2D systems, which illustrates how the local-density functionals
obtained in this work may be used to describe non-uniform 
electronic systems in a strong magnetic field.
The results presented here may be of use in a wide variety
of problems of current physical interest.
In particular, our results for 2D systems should be of relevance to
researchers interested in density-functional methods for low-dimensional 
electronic systems.

Although the present work has focused on spin-averaged densities at zero
temperature, an extension to include spin and finite temperatures is
straightforward.  

\section{acknowledgements}
It is a pleasure to thank Dr. R. K. Bhaduri for stimulating discussions.
The hospitality of the Department of Physics and Astronomy at 
McMaster University, Hamilton, Canada, is also gratefully acknowledged.  
Financial support for this work was provided by the NSERC of Canada and
Queen's University, Kingston, Canada.


\begin{references}
\bibitem{note1}
The choice of gauge is arbitrary, because the total energy-functional
must be gauge invariant.
\bibitem{note2}
Of course, care must be still be exercised at the branch point, $s=0$, when
evaluating the large $\gamma$ behaviour of ${\cal F}^{(d)}(\gamma)$.
\end{references}
\end{document}